\documentclass[10pt,conference]{IEEEtran}

\usepackage{enumerate}

\usepackage{epsfig}

\usepackage{amssymb}
\usepackage{amsmath}
\usepackage{latexsym}

\newtheorem{theorem}{Theorem}

\newtheorem{lemma}{Lemma}

\newtheorem{definition}{Definition}

\newcommand{\mX}{\mathcal X}

\newcommand{\mC}{\mathcal C}
\newcommand{\mE}{\mathcal E}

\newcommand{\mW}{\mathcal W}

\newcommand{\ve}{\varepsilon}

\newcommand{\C}[2]{{#1}\wedge{#2}}

\newcommand{\cancel}[1]{}

\DeclareMathOperator{\comp}{comp}
\DeclareMathOperator{\ext}{ext}
\DeclareMathOperator{\comm}{comm}
\DeclareMathOperator{\asym}{asym}

\DeclareMathOperator{\prob}{Pr}

\date{}

\author{
\authorblockN{
Renato Renner\authorrefmark{1}\qquad
Stefan Wolf\authorrefmark{2}\qquad
J\"urg Wullschleger\authorrefmark{2}
}
\authorblockA{
\authorrefmark{1}Centre for Quantum Computation, University of Cambridge, United Kingdom\\
E-mail: r.renner@damtp.cam.ac.uk}
\authorrefmark{2}Computer Science Department,
ETH Z\"urich,
Switzerland\\
E-mail: \{wolf,wjuerg\}@inf.ethz.ch}

\title{The Single-Serving Channel Capacity}

\begin{document}

\maketitle
\begin{abstract}
  In this paper we provide the answer to the following question:
  Given a noisy channel $P_{Y|X}$ and $\varepsilon>0$, how many bits
  can be transmitted with an error of at most $\varepsilon$ by a
  single use of the channel?
\end{abstract}

\section{Introduction}

Shannon entropy and information~\cite{shannon48} have been shown very 
significant in the scenario of i.i.d. distributions and asymptotic rates. 
Unfortunately, however, these two assumptions fail to be realistic 
in many real-world scenarios. First of all, a given primitive 
or random experiment is actually available only a limited number 
of times, and an asymptotic analysis has, therefore, a limited 
significance. Second, the assumption that a certain primitive 
is repeated {\em independently\/} many times is not always 
realistic. An important example is cryptography, where this
assumption leads to a strong restriction on the adversary's
behavior and possibilities. 

In~\cite{HanVer93}, the assumption of independence has been dropped, but 
the analysis still remains asymptotic. In the present paper, we 
drop {\em both\/} assumptions at once and consider the case where a
certain information-theoretic primitive, such as a communication channel, 
or random experiment is available only {\em once}. This {\em 
single-serving\/} case has also been called "single shot" in the literature. 

Let us first consider an example from cryptography or, more precisely, 
information-theoretic key agreement from correlated pieces of 
information. Let two parties, Alice and Bob, as well as
an adversary, Eve, have access to $n$ \emph{independent realizations}
of random variables $X$, $Y$, and $Z$, respectively, with joint
probability distribution $P_{X Y Z}$.  Moreover, authenticated but
public communication from Alice to Bob (but not in the other
direction) is possible. Their goal is to generate a common secret key
of length $\ell(n)$, i.e., a uniform string about which the adversary
is virtually ignorant. Asymptotically, for large $n$, the \emph{rate}
at which such a key can be generated is given by
\begin{equation}
\lim_{n \to \infty} \frac{\ell(n)}{n}
= 
\max_{YZ\leftrightarrow X\leftrightarrow UV}(H(U|ZV)-H(U|YV))\label{eins}
\end{equation}
(see, for instance, \cite{Wyner48,CisKoe78,AhlCis93,Maurer93a}).

Let us now consider the \emph{non-asymptotic} case where $n=1$, i.e.,
the random experiment defined by $P_{X Y Z}$ is only run \emph{once}.
How many virtually secret bits can then be extracted?  First of all,
note that \eqref{eins} fails to provide the correct answer in this
case. To see this, assume, e.g., that $X$ is uniformly distributed and
that $Y=X$, whereas $Z=X$ holds with probability $1/2$ (and $Z=\Delta$
otherwise).  Then
the right-hand side of 
 \eqref{eins} is non-zero, but no secret can be
extracted at all by Alice and Bob since, with probability $1/2$, Eve
knows everything. We conclude that Shannon entropy fails to be the
right measure in this setting. But what does it have to be replaced
by?

Results on {\em randomness extraction\/}, also known as
{\em privacy amplification\/}~\cite{BeBrRo88,ILL89,HILL99}, indicate that the right
answer might be given by so-called \emph{min-entropies} rather than
Shannon entropies.  Indeed, it is shown in~\cite{RenWol05} that the
so-called {\em conditional smooth min- and
  max-entropies\/}~\cite{RenWol04a, RenWol05} $H_{\max}^\varepsilon$
and $H_{\min}^\varepsilon$ (for the precise definitions see below)
replace Shannon entropy in this case; the achievable secret-key length
$\ell$ is approximated (up to a term $\log(1/\varepsilon)$,
where $\varepsilon$ is the security of the final key) by
\[
\ell \approx \max_{YZ\leftrightarrow X\leftrightarrow UV}(H_{\min}^\varepsilon(U|ZV)-H_{\max}^\varepsilon(U|YV))\ .
\]

It is the goal of this paper to show that smooth min- and max-entropy
has a similar significance in communication theory, i.e., it can be
used for the characterization of communication tasks in a single-serving
setting. Among others, we consider the following question: Given a
noisy communication channel $\mW = P_{Y|X}$ and $\varepsilon>0$, what
is the maximum number $C_{\comm}^{\varepsilon}(\mW)$ of bits that can
be transmitted with error at most $\varepsilon$ by {\em a single
  use\/} of the channel. Recall that, in the i.i.d.\ case, i.e., if
the channel can be used many times independently, an asymptotic answer
to this question is given by the channel capacity $C_{\comm}^{\asym}$,
which can be expressed by the well-known formula \cite{shannon48}
\[
C_{\comm}^{\asym}(\mW) = \max_{P_X}(H(X)-H(X|Y)) \ .
\]
As we shall see, the answer for the single-serving case looks very
similar, but the (conditional) Shannon entropies are replaced by
smooth min- and max-entropies:
\[
C_{\comm}^{\varepsilon}(\mW) \approx \max_{P_X}(H_{\min}^{\varepsilon}(X)-H_{\max}^{\varepsilon}(X|Y))\ .
\]
%
%
\section{Notation and Previous Work}

\subsection{Smooth Min- and Max-Entropies}

Let $X$ be a random variable with probability distribution $P_X$. The
\emph{max-entropy} of $X$ is defined as the binary logarithm of the
size of the support of $P_X$, i.e.,
\begin{align*}
H_{\max}(X) &= \log |\{x \in \mX: P_X(x) > 0\}| \ .
\end{align*}
Similarly, the \emph{min-entropy} of $X$ is given by the negative
logarithm of the maximum probability of $P_X$, i.e.,
\begin{align*}
 H_{\min}(X) &= -\log(\max_x(P_X(x))\ .
\end{align*}
Note that $H_{\min}(X) \leq H(X) \leq H_{\max}(X)$, i.e., the min- and
max-entropies are lower and upper bounds for the Shannon entropy (and
also for any R\'enyi entropy of order $\alpha \in [0,\infty]$),
respectively.

For random variables $X$ and $Y$ with joint
distribution $P_{X Y}$, the ``conditional'' versions of these entropic quantities are
defined as
\begin{align*}
H_{\max}(X|Y) &= \max_{y} H_{\max}(X|Y=y)\ , \\
H_{\min}(X|Y) &= \min_{y} H_{\min}(X|Y=y)\ .
\end{align*}
In \cite{RenWol05}, max- and min-entropies have been generalized to
so-called \emph{smooth max- and min-entropies}. For any $\varepsilon
\geq 0$, they are defined by optimizing the "non-smooth" quantities
over all random variables $\bar{X}$ and $\bar{Y}$ which are equal to
$X$ and $Y$ except with probability $\varepsilon$, i.e.,
\begin{align*}
H^\ve_{\max}(X|Y) &= \min_{\bar X \bar Y: \prob[XY \neq \bar X \bar Y] \leq \ve} H_{\max}(\bar X|\bar Y) \\
H^\ve_{\min}(X|Y) &= \max_{\bar X \bar Y: \prob[XY \neq \bar X \bar Y] \leq \ve} H_{\min}(\bar X|\bar Y) .
\end{align*}
Equivalently, smooth max- and min-entropies can be expressed in terms
of a optimization over events $\mE$ that have probability at least
$1-\ve$. Let $P_{X \mE|Y=y}(x)$ be the probability that $X=x$
\emph{and} the event $\mE$ occurs, conditioned on $Y=y$. We then have
\begin{align*}
  H^\ve_{\max}(X|Y) &= \min_{\mE: \Pr(\mE)\geq 1-\ve} \max_{y} \log |\{x: P_{X \mE|Y=y}(x) > 0\}|
\\
  H^\ve_{\min}(X|Y) &= \max_{\mE: \Pr(\mE)\geq 1-\ve} \min_{y} \min_x (-\log P_{X \mE|Y=y}(x)) .
\end{align*}

These smooth entropies 
have properties similar to Shannon entropy|this is 
in contrast the the
usual, non-smooth min- and max-entropies which have many counterintuitive
properties that make them less useful in many contexts.
For example, the
\emph{chain rule} $H(X|Y) = H(XY)-H(Y)$ translates to~\cite{RenWol05}
\begin{align*}
& H^{\ve+\ve'}_{\max}(XY) - H^{\ve'}_{\max}(Y) \leq H^{\ve}_{\max}(X|Y)\ , \\
&\qquad \leq H^{\ve_1}_{\max}(XY) - H^{\ve_2}_{\min}(Y) + \log(1/(\ve-\ve_1-\ve_2))
\end{align*}
and
\begin{align*}
& H^{\ve_1}_{\min}(XY) - H^{\ve_2}_{\max}(Y) - \log(1/(\ve-\ve_1-\ve_2)) \\
&\qquad \leq H^\ve_{\min}(X|Y) \leq H^{\ve+\ve'}_{\min}(XY) - H^{\ve'}_{\min}(Y).
\end{align*}

\subsection{Operational Interpretation of Smooth Max- and Min-Entropies}
In \cite{SleWol73} it was shown 
that the rate at which many independent realizations of $X$ can be compressed
is asymptotically $H(X|Y)$ if the decoder is provided with side information
$Y$. It is easy to see that $H(X|Y)$ also is the rate
at which uniform randomness can be extracted from $X$, in such a way that it is independent
of $Y$.
In \cite{RenWol05}, it was shown that the smooth entropies
$H^\ve_{\max}$ and $H^\ve_{\min}$ quantify compression and randomness
extraction, respectively, in the single-serving case.
More precisely, let $H^\ve_{\comp}(X|Y)$ be
the length of a bit string needed to store \emph{one} instance of $X$
such that $X$ can later be recovered with an error of at most $\ve$
using this string and $Y$. This quantity is then roughly equal to
$H^\ve_{\max}$, i.e.,
\begin{align*}
& H^\ve_{\max}(X|Y)
\leq H^\ve_{\comp} (X|Y) \\
&\qquad \leq H^{\ve'}_{\max}(X|Y) + \log(1/(\ve - \ve')) \ .
\end{align*}
Similarly, let $H^\ve_{\ext}(X|Y)$ be the maximum length of a string
that can be computed from $X$, such that this string is uniformly
distributed and independent of $Y$, with an error of at most $\ve$. We
then have
\begin{align*}
& H^{\ve'}_{\min}(X|Y) - 2 \log(1/(\ve - \ve')) \\
&\qquad \leq H^\ve_{\ext}(X|Y)
\leq H^{\ve}_{\min}(X|Y).
\end{align*}

\subsection{Common Information}

The \emph{common information} is the 
rate at which 
 uniform random bits can be extracted both from $X^n$ and $Y^n$,
 which come from independent
 repeated realizations of the random experiment $P_{XY}$ 
without communicating. It has
been shown in \cite{GacKoe73} that the common information is equal to
the maximum entropy of a common random variable that
both players can compute. As in \cite{FiWoWu04, WolWul04a}, we will
denote this random variable by $\C{X}{Y}$, i.e., the common
information of $X$ and $Y$ is given by $H(\C{X}{Y})$.

It is shown in \cite{WolWul04a} that the common information can be
used to characterize the zero-error capacity $C^{\asym}_{0\textrm{-}\comm}(\mW)$ of a channel
$\mW$ as follows:
\[
C^{\asym}_{0\textrm{-}\comm}(\mW) = \lim_{n \rightarrow \infty} \max_{P_{X^n}} \frac 1 n H(\C{X^n}{Y^n})\ .
\]
Note that the usual (asymptotic) channel capacity
$C_{\comm}^{\asym}(\mW)$ of $\mW$ is given by a similar
expression, where the common information is replaced by the mutual
information, i.e.,
\[C_{\comm}^{\asym}(\mW) = \max_{P_{X}}I(X;Y) = \lim_{n\rightarrow \infty} \max_{P_{X^n}} \frac 1 n I(X^n;Y^n)\ .\]

\section{Extractable Common Randomness}

We denote by $C^{\ve}_{\ext}(X,Y)$ the maximum amount of uniform
randomness that can be extracted from $X$ and $Y$, without any
communication, with an error of at most $\ve$.  Asymptotically, it
follows from \cite{GacKoe73} that
\[
\lim_{\ve \rightarrow 0}
\lim_{n \rightarrow \infty}
\frac {C^{\ve}_{\ext}(X^n,Y^n)}{n}
= H(\C{X}{Y}).
\]

In the following, we analyze the quantity $C^{\ve}_{\ext}(X,Y)$ for the
single-serving case.  First, we will show that $C^{\ve}_{\ext}(X,Y)$ is
characterized by the following quantity.
\begin{definition}
\[ C^{\ve}_{\min}(X;Y) = \max_{\bar X \bar Y: \Pr[\bar X \bar Y \neq X Y] \leq \ve} H_{\min}(\C{\bar X}{\bar Y})\ . \]
\end{definition}

\begin{theorem} 
  For all random variables $X$ and $Y$, and for all $\ve'$ and $\ve >
  \ve'$, we have
  \[
  C^{\ve}_{\ext}(X;Y) \geq C^{\ve'}_{\min}(X;Y) - 2 \log(1/(\ve-\ve')) \ .
  \]
\end{theorem}

\begin{proof}
  Let Alice and Bob have $\bar X$ and $\bar Y$, respectively.  They
  both can calculate $\C{\bar X}{\bar Y}$ and extract at least
  $H_{\min}(\C{\bar X}{\bar Y}) - 2 \log(1/(\ve - \ve'))$ bits with an
  error of at most ${\ve - \ve'}$.  Since $\Pr[\bar X \bar Y \neq X Y]
  \leq \ve'$, we get at most an additional error of $\ve'$ if they use
  $X$ and $Y$ instead of $\bar X$ and $\bar Y$. The total error is,
  therefore, at most $\ve$.
\end{proof}

\begin{theorem} 
  For all random variables $X$ and $Y$, and for all $\ve$, we have
\[C^{\ve}_{\ext}(X;Y) \leq C^{\ve}_{\min}(X;Y)\ .\]
\end{theorem}

\begin{proof}
  Let us assume that Alice and Bob can extract more than
  $C^{\ve}_{\min}(X;Y)$ bits with an error at most $\ve$. Therefore
  there exist functions $f$ and $g$ such that with probability $1 -
  \ve$ both functions output the same uniform random string $R$ of
  length bigger than $C^{\ve}_{\min}(X;Y)$, which means that there
  exist $\bar X, \bar Y$ such that $\prob[(\bar X, \bar Y) \neq (X,
  Y)] \leq \ve$ and $f(\bar X) = g(\bar Y) = R$.  As shown in Lemma~1
  of~\cite{WolWul04a}, this implies that $R$ can be computed from
  $\C{\bar X}{\bar Y}$, that is, there exists a function $h$ such that
  $R = h(\C{\bar X}{\bar Y})$. The function $h$ could thus be used to
  extract more than $H_{\min}$ bit from $\C{\bar X}{\bar Y}$, which is
  impossible.
\end{proof}

In the following, we derive an upper bound on $C^{\ve}_{\min}(X;Y)$ in
terms of smooth min- and max-entropies.


\begin{lemma} \label{C-0}
For all random variables $X$ and $Y$, and for all $\ve$, $\ve_1$, and $\ve_2$, we have
\[ C^{\ve}_{\min}(X;Y) \leq H^{\ve_2}_{\max}(X) - H^{\ve_1 + \ve_2 + 2 \ve}_{\max}(X|Y) + \log(1/\ve_1)\ .\]
\end{lemma}

\begin{proof}
  Let $\bar X$ and $\bar Y$ be the random variables that maximize
  $C^{\ve}_{\min}(X;Y)$, and let $C = \C{\bar X}{\bar Y}$.  We have
\[ H_{\min}(C) \leq H^{\ve}_{\max}(XC) - H^{\ve_1 + \ve_2}_{\max}(X|C) + \log(1/\ve_1). \]
$C$ is a function of $X$ and of $Y$ with probability at least $1-\ve$.
Therefore, we can bound
\[H^{\ve_2}_{\max}(XC) \leq H^{\ve_2 - \ve}_{\max}(X) \]
and
\[H^{\ve_1 + \ve_2}_{\max}(X|C) \geq H^{\ve_1 + \ve_2 + \ve}_{\max}(X|Y). \]
We get
\[ H_{\min}(C) \leq H^{\ve_2 - \ve}_{\max}(X) - H^{\ve_1 + \ve_2 + \ve}_{\max}(X|Y) + \log(1/\ve_1). \]
The statement follows when $\ve$ is added to $\ve_2$.
\end{proof}

No non-trivial lower bound is known so far for $C^{\ve}_{\ext}(X,Y)$.
However, one can bound $\max_{P_X} C^{\ve}_{\min}(X;Y)$. This will
turn out to be useful for the considerations in the next section.

\begin{lemma}
  For all conditional distributions $P_{Y|X}$ and for all $\ve_1$, $\ve_2$,
  and $\ve_3$, we have
\begin{align*}
  & \max_{P_X} C^{\ve_1 + \ve_2 + \ve_3}_{\min}(X;Y) \\
  &\qquad \geq \max_{P_X} \left( H^{\ve_1}_{\min}(X) -
    H^{\ve_2}_{\max}(X|Y) \right ) - \log(1/\ve_3)\ .\ \ \ 
\end{align*}
\end{lemma}

\begin{proof}
  Let $P_X$ be the distribution that maximizes $H^{\ve_1}_{\min}(X) -
  H^{\ve_2}_{\max}(X|Y)$. There exist random variables $\bar X$ and $\bar
  Y$ with $\prob[XY \neq \bar X \bar Y] \leq \ve_1 + \ve_2$ such that
  $H_{\min}(\bar X) - H_{\max}(\bar X|\bar Y) = H^{\ve_1}_{\min}(X) -
  H^{\ve_2}_{\max}(X|Y)$.  We choose, independently and according to the
  distribution $P_{\bar X}$, $2^{H_{\min}(\bar X) - H_{\max}(\bar X|\bar Y)
    - \log(1/\ve_3)}$ values. Let $S$ be the set of these values and
  let $\tilde X$ be a random variable that takes on a value in $S$
  with equal probability.  Since $P_{\bar X}(x)\cdot2^{H_{\min}(\bar
    X)} \leq 1$, the probability that a value $x$ chosen according to
  $P_{\bar X}$ is in $S$ is at most
\[
P_{\bar X}(x)\cdot2^{H_{\min}(\bar X) - H_{\max}(\bar X|\bar Y) - \log(1/\ve_3)}
\leq 2^{- H_{\max}(\bar X|\bar Y)} \ve_3\ .
\]
Let $\tilde x$ and $\tilde y$ be chosen according to the distribution
$P_{\tilde X}P_{\bar Y|\bar X}$. The probability that there exists a
value $\tilde x' \in S$ such that $\tilde x' \neq \tilde x$ and
$P_{\bar Y | \bar X}(\tilde y,\tilde x') > 0$ is at most $2^{H_{\max}(\bar
  X|\bar Y)} 2^{- H_{\max}(\bar X|\bar Y)}\ve_3 = \ve_3$.  Therefore, there
exists a function $f$ such that $\Pr[ \tilde X \neq f(\tilde Y)] \leq
\ve_3$ holds, and we have
\begin{align*}
C^{\ve_3}_{\min}(\bar X; \bar Y)
&= H_{\min}(\tilde X) \\
&= H^{\ve_1}_{\min}(X) - H^{\ve_2}_{\max}(X|Y) - \log(1/\ve_3)\ .
\end{align*}
The statement now follows from the fact that
\[C^{\ve_1 + \ve_2 + \ve_3}_{\min}(X; Y) \geq C^{\ve_3}_{\min}(\bar X; \bar Y)\ .\]
\end{proof}

\section{Communication}

Let us now come back to the question posed in the abstract. We define
the \emph{$\ve$ single-serving channel capacity} of a channel $\mW =
P_{Y|X}$, denoted $C^{\ve}_{\comm}(\mW)$, as the maximum number of bits
(i.e., the logarithm of the number of symbols) that can be transmitted
in {\em a single use\/} of $\mW $, such that every symbol can be decoded
by an error of at most $\ve$.
Theorem~\ref{main1} shows the connection between the the extractable
common randomness and single-serving channel capacity, similar to the
connection between the common information and the
zero-error capacity shown in \cite{WolWul04a}.
\begin{theorem} \label{main1}
For all channels $\mW = P_{Y|X}$ and for $\ve' < \ve$ , we have
\begin{align*}
&\max_{P_X} C^{\ve'}_{\min}(X;Y) - \log(\ve/(\ve-\ve')) \\
&\qquad \qquad \leq C^{\ve}_{\comm}(\mW) \leq \max_{P_X} C^{\ve}_{\min}(X;Y)\ .
\end{align*}
\end{theorem}

\begin{proof}
Let $\mC \subset \mX$ be a code that can be decoded with an error of at most $\ve$
and let $X$ be uniformly distributed over $\mC$. Then there exists a $\bar Y$ with
$\Pr[\bar Y=Y] \geq 1-\ve$, such that $X = \C{X}{\bar Y}$. It follows that
\[ \max_{P_X} C^{\ve}_{\min}(X;Y) \geq C^{\ve}_{\comm}(\mW)\ . \]
Let $P_X$ be a distribution that maximizes $\max_{P_X} C^{\ve'}_{\min}(X;Y)$, and let
$\bar X, \bar Y$ be random variables 
for which $H(\bar{X}\wedge\bar{Y})=C^{\ve'}_{\min}(X;Y)$ holds as well as 
 $\Pr[\bar X \bar Y = XY] \geq 1- \ve'$. Let $C := \C{\bar X}{\bar Y}$.
We can write $C$ as a combination of uniform random variables $C_i$, with
$H_{\min}(C_i) = H_{\min}(C)$.
More precisely, we have $P_C=\sum_i{\lambda_i P_{C_i}}$, where $P_{C_i}(x)\in\{0,2^{-H_{\min}(C)}\}$ for all $x$.  
 The support of 
the 
 random variable $C_i$ which minimizes the error probability defines a
code $\mC_i \subset \mX$ that can be
decoded with an error of at most $\ve$, if the input is uniformly distributed.
Since we need a code that works for \emph{any} input distribution, we delete all symbols 
which get decoded with an error bigger than $\ve > \ve'$. From the Markov inequality
follows that the reduced code still contains at least $\frac{\ve - \ve'}{\ve} 2^{H_{\min}(C)}$ symbols.
It follows that
\[ C^{\ve}_{\comm}(\mW) \geq \max_{P_X} C^{\ve'}_{\min}(X;Y) - \log(\ve/(\ve-\ve'))\ . \]
\end{proof}

From Lemma \ref{C-0}  we have
\begin{align*}
& \max_{P_X}  C^{\ve}_{\min}(X;Y) \\
&\qquad \leq \max_{P_X} \left(H^{\ve_2}_{\max}(X) - H^{\ve_1 + \ve_2 + 2 \ve}_{\max}(X|Y)\right ) + \log \frac 1 {\ve_1}.
\end{align*}
From the same argument as in the proof of Theorem \ref{main1} follows that
$\max_{P_X}  C^{\ve}_{\min}(X;Y)$ is maximized by a distribution
where all $x$ with positive probability have equal probabilities.
Therefore, we have $H^{\ve}_{\max}(X) = H^{\ve}_{\min}(X)$ and get
\begin{align*}
& \max_{P_X} \left( H^{\ve'}_{\min}(X) - H^{\ve''}_{\max}(X|Y) \right ) - \log \frac 1 {\ve - \ve' - \ve''} \\
&\qquad \leq \max_{P_X} C^{\ve}_{\min}(X;Y) \\
&\qquad \leq \max_{P_X} \left ( H^{\ve_2}_{\min}(X) - H^{\ve_1 + \ve_2 + 2 \ve}_{\max}(X|Y) \right ) + \log \frac 1 {\ve_1}\ .
\end{align*}

Together with Theorem~\ref{main1}, this implies the following bound on
the single-serving channel capacity $C^{\ve}_{\comm}(\mW)$.

\begin{theorem} \label{thm:mainchannel} For all channels $\mW = P_{Y|X}$ and all $\ve'$, $\ve''$, $\ve > \ve' + \ve''$, $\ve_1$, and $\ve_2$, we have
\begin{align*}
& \max_{P_X} \left( H^{\ve'}_{\min}(X) - H^{\ve''}_{\max}(X|Y) \right ) - \log \frac {4\ve} {(\ve - \ve' - \ve'')^2} \\
&\qquad \leq C^{\ve}_{\comm}(\mW) \\
&\qquad \leq 
\max_{P_X} \left( H^{\ve_2}_{\min}(X) - H^{\ve_1 + \ve_2 + 2 \ve}_{\max}(X|Y) \right ) + \log \frac 1 {\ve_1}\ .
\end{align*}
\end{theorem}
%

\section{Conclusions}
Shannon entropy can be used to characterize a variety of
information-processing tasks such as communication over noisy
channels in the scenario where the primitive can be 
used independently many times. We have shown that smooth min- and max-entropies play a
similar role in the more general single-serving case. In particular, we
have given an explicit expression for the ``single-serving channel
capacity.'' We suggest as an open problem to find other such examples
and contexts.

The notion of conditional smooth entropies has recently been
generalized to quantum information theory~\cite{Renner05} (see
also~\cite{HayWin03} for the non-conditional case). It is likely (but
still
unproven) that, similarly to our classical
Theorem~\ref{thm:mainchannel}, these quantities can be used to
characterize single-serving capacities of quantum channels.

\newpage

\section*{Acknowledgment}

This work was supported by the Swiss National Science Foundation (SNF)
and Hewlett Packard Research Labs.


\end{document}